\documentclass[a4paper, 10pt, conference]{IEEEtran}
\IEEEoverridecommandlockouts
\usepackage{cite}
\usepackage{amsmath,amssymb,amsfonts}
\usepackage{algorithmic}
\usepackage{graphicx}
\usepackage{textcomp}
\usepackage{xcolor}
\def\BibTeX{{\rm B\kern-.05em{\sc i\kern-.025em b}\kern-.08em
    T\kern-.1667em\lower.7ex\hbox{E}\kern-.125emX}}
\begin{document}

\title{Spatial encoding of BOLD fMRI time series for categorizing static images across visual datasets: A pilot study on human vision\\
\thanks{This paper is accepted for publication in IEEE Region 10 Technical conference, TENCON 2023, to be held in Chiang Mai, Thailand from 31 October - 3 November, 2023.}
}

\author{\IEEEauthorblockN{1\textsuperscript{st} Vamshi K. Kancharala}
\IEEEauthorblockA
{\textit{Networking and Communication} \\
\textit{IIIT Bangalore}\\
Bangalore, India \\
kancharla.vamshi@iiitb.ac.in}
\and
\IEEEauthorblockN{2\textsuperscript{nd} Debanjali Bhattacharya}
\IEEEauthorblockA{\textit{Networking and Communication} \\
\textit{IIIT Bangalore}\\
Bangalore, India \\
debanjali.bhattacharya@iiitb.ac.in}
\and
\IEEEauthorblockN{3\textsuperscript{rd} Neelam Sinha}
\IEEEauthorblockA{\textit{Networking and Communication} \\
\textit{IIIT Bangalore}\\
Bangalore, India \\
neelam.sinha@iiitb.ac.in}
}

\maketitle

\begin{abstract}

Functional MRI (fMRI) is widely used to examine brain functionality by detecting alteration in oxygenated blood flow that arises with brain activity. In this study, complexity-specific image categorization across different visual datasets is performed using fMRI time series (TS) to understand differences in neuronal activities related to vision. Publicly available BOLD5000 dataset is used for this purpose, containing fMRI scans while viewing 5254 images of diverse categories, drawn from three standard computer vision datasets: COCO, ImageNet and SUN. To understand vision, it is important to study how brain functions while looking at different images. To achieve this, spatial encoding of fMRI BOLD TS has been performed that uses classical Gramian Angular Field (GAF) and Markov Transition Field (MTF) to obtain 2D BOLD TS, representing images of COCO, Imagenet and SUN. For classification, individual GAF and MTF features are fed into regular CNN. Subsequently, parallel CNN model is employed that uses combined 2D features for classifying images across COCO, Imagenet and SUN. The result of 2D CNN models is also compared with 1D LSTM and Bi-LSTM that utilizes raw fMRI BOLD signal for classification. It is seen that parallel CNN model outperforms other network models with an improvement of 7\% for multi-class classification.
\newline
\indent \textit{Clinical relevance}— The obtained result of this analysis establishes a baseline in studying how differently human brain functions while looking at images of diverse complexities.
\end{abstract}

\section{Introduction}
\label{sec:intro}
Studies on neuronal activities of human brain have increased remarkably in past decades in order to understand the complex characteristics of human brain underlying cognition, behavior, and perception. For this, functional Magnetic Resonance Imaging (fMRI) is widely used  that exploits the changes in magnetic properties of oxygenated and de-oxygenated blood to measure the fluctuation in neuronal activities during well-defined tasks or resting-state condition. During over last two decades several BOLD fMRI studies have been attempted to highlight changes in the activation of brain areas and for dysfunctions detection in patients with neurological and psychiatric disorders. However, studies on human visual perception has been still limited due to complex experimental procedures involved in generating adequate number of good quality fMRI neuro-image data, representing vision. The release of publicly available dataset “BOLD5000” \cite{b1}, that contains fMRI scans of subjects acquired while viewing 5254 images, has made it possible to study the brain dynamics during visual tasks in greater detail. These images are chosen from three well-known computer vision datasets: (i) COCO, (ii) ImageNet, and (iii) SUN, that contain images with diverse complexities for performing different computer vision related tasks. There are few studies reported in literature that used BOLD5000 dataset for tasks like classification of well-separable image classes \cite{b2}, for pre-training to predict cognitive fatigue in traumatic brain injury \cite{b3} and for neural encoding\cite{b4}.\par
Different from these studies, in the current work, we have utilized the 1D and 2D representation of BOLD fMRI time series (TS) data to categorize diverse image complexities across these visual datasets. The main objective of the study is to hypothesize the fact that \textit{BOLD activities in brain are different when we look at images having varying complexity, object-scales and context}. To achieve this, 1D fMRI TS is encoded into its 2D representation using Gramian Angular Field (GAF) and Markov Transition Field (MTF) in order to capture the spatial information of BOLD TS. The GAF and MTF are well-known TS encoding techniques \cite{b5,b6}, having its application in different areas like, performing rolling bearing fault diagnosis and classification using vibration sensor data \cite{b7,b8}, EEG signal classification \cite{b9,b10}, single residential load forecasting \cite{b11}, manufacturing quality prediction \cite{b12}, milling tool condition monitoring \cite{b13}, malware classification \cite{b14} and fault classification of marine systems \cite{b15}.\par 
The main contribution of this work is to investigate the efficacy of 2D GAF and MTF as obtained from fMRI BOLD TS of active voxels during visual task, for the purpose of categorizing visual stimuli of diverse image complexities into corresponding datasets of Imagenet, COCO and SUN. 

\begin{figure}[htbp]
\centering
\begin{minipage}[b]{0.3\linewidth}
  \centering
  \centerline{\includegraphics[width=2.6cm]{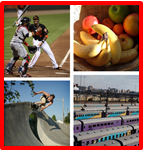}}
  \centerline{(a)}\medskip
\end{minipage}
\begin{minipage}[b]{0.3\linewidth}
  \centering
  \centerline{\includegraphics[width=2.7cm]{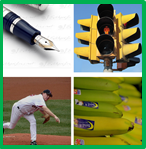}}
  \centerline{(b)}\medskip
\end{minipage}
\begin{minipage}[b]{0.32\linewidth}
  \centering
  \centerline{\includegraphics[width=2.75cm]{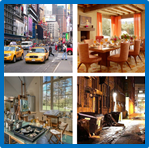}}
  \centerline{(c)}\medskip
\end{minipage}
\caption{Sample images, taken from the three computer vision datasets having different complexities: (a) COCO (\textit{Red box}): contains multiple objects and actions, (b) ImageNet (\textit{Green box}): contains single-focused object and (c) SUN (\textit{Blue box}): contains indoor and outdoor scenes.}
\label{fig:imgs}
\end{figure}

\section{Dataset description}
\label{sec:dataset}
The publicly available BOLD5000 dataset is used for this study \cite{b1}. Four right-handed healthy volunteers (M:F 1:3, age: 24-27 years) were selected from Carnegie Mellon University. FMRI scans are acquired from these participants while viewing 5254 images of diverse categories. These images were selected from three classical computer vision database: \\
(i) \textit{Common objects in context (COCO)}, which is a standard benchmark dataset comprising of images of complex indoor and outdoor scenes. 2000 images from COCO dataset were used as visual stimuli. The multiple objects in COCO dataset depict interaction with other objects in realistic context.\\
(ii) \textit{Scene Understanding (SUN)} dataset that contains real-world scene images of indoor/outdoor environment. The set of 1,000 scene images covering 250 categories were selected which inclined to be more panoramic, having no focus on specific object. \\
(iii) \textit{ImageNet} dataset, which contains singular-focused object in real-world scenes. 1916 images were selected from ImageNet that mainly focus on a single object. Unlike COCO and SUN images, the single object in the ImageNet images is mostly placed at center and occupy a large fraction of image, so that it is distinguishable clearly from its background. The sample images from COCO, ImageNet and SUN are shown in Figure~\ref{fig:imgs}. \par
Each functional session consisted of 9 to 10 runs where in each run 37 stimuli (images) were presented randomly to the participants. The fMRI data was acquired using a 3T Siemens Verio MR scanner using a 32-channel phased array head coil. Further details on subject demographics, stimuli selection, fMRI scan acquisition and data pre-processing procedures can be found in \cite{b1}.

\section{Proposed methodology}
\begin{figure*}[htbp]
\centering
\includegraphics[width=16cm]{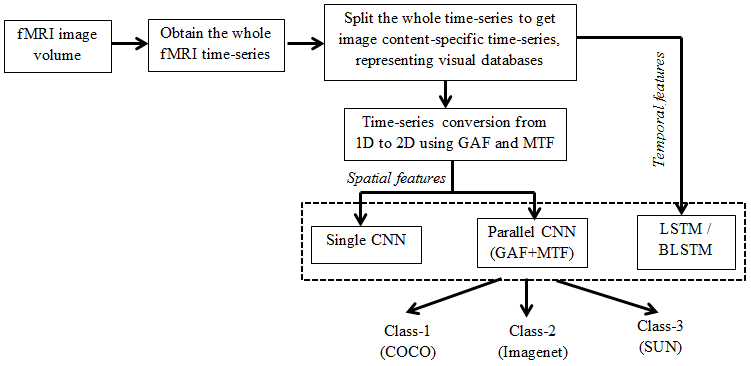} 
\caption{Block schematic of the proposed methodology}
\label{fig:bd}
\end{figure*}

The block schematic of the proposed methodology is shown in Figure~\ref{fig:bd}. The whole TS is extracted from each active voxel in each fMRI trail. For categorizing images into corresponding visual datasets (ImageNet, COCO and SUN), the whole TS is split into three parts according to the images viewed by the participants from these three datasets, during each fMRI trail. These voxel-specific TS are then used to train neural networks for image categorization across visual datasets. The codes of this experiment are available at the following link. \footnote{https://github.com/kancharlavamshi/Spatial-encoding-of-BOLD-fmri-time-series-for-categorical-static-images-across-visual-dataset}

\subsection{FMRI Time Series Extraction}
\label{ref:FMRI Time Series Extraction}

\begin{figure}[htbp]
\centering
\includegraphics[width=8cm]{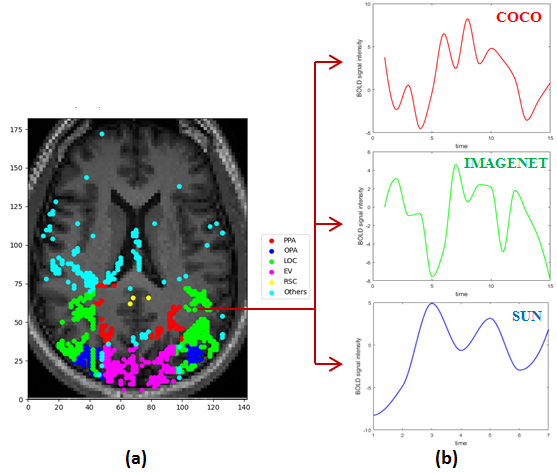} 
\caption{fMRI Time series extraction from active voxel location in brain. \textbf{(a)} Mapping of the location of voxel activation on fMRI image for a representative subject. Voxel activation at different visual ROIs (as defined in \cite{b1}) are shown by different colors. \textbf{(b)} it shows the extracted time series that represents 3 distinct datasets for a specific active voxel (top: COCO, middle: Imagenet, bottom: SUN).}
\label{fig:TimeSeries}
\end{figure}

In order to extract BOLD fMRI TS, it is necessary to have the information about active voxels locations in brain while viewing the images. In our study, the well-known SPM toolbox is utilized to get active voxel locations from 4D \textit{(x,y,z,t)} fMRI data. The locations of active voxels are mostly found within 5 visual ROIs as defined in \cite{b1}, which are the parahippocampal place area (PPA), the retrosplenial complex (RSC), the occipital place area (OPA), Early visual area (EV) and lateral occipital complex (LOC). Few voxels which are found to be active in other sub-cortical regions, labeled as "others" in this study (Figure~\ref{fig:TimeSeries}: \textit{left}). 
From each fMRI trail and for each active voxel, the whole TS which is the representation of BOLD intensity distribution over time, was extracted. Further, in order to obtain the image complexity-specific fMRI TS, the whole TS of length 37 (since, the no. of stimuli = 37, at each trial) is separated into three parts as shown in Figure~\ref{fig:TimeSeries} (\textit{right}). Each of these three TS illustrates the BOLD intensity distribution while viewing images from COCO (\textit{Red}), ImageNet (\textit{Green}) and SUN (\textit{Blue}). This voxel-specific TS is detrened and Z-score normalized before it is fed to deep neural network model. 

\subsection{Spatial Encoding of fMRI Time Series Signal}

The fMRI TS are converted into its 2D image representation using Gramian Angular Field (GAF) and Markov Transition Field (MTF) \cite{b5,b6} in order to produce spatial features and visual patterns which are not apparent in the 1D TS, without disturbing the temporal dependency of the fMRI TS. The mapping of 2D spatial encoding of fMRI TS is done with expectation that, this may help deep neural networks to learn more effectively. 

\subsubsection{Gramian Angular Field}
Let us consider fMRI time series as $Z=\{z_{1}, z_{2}, z_{3}, \dots, z_{n}\}$. Min-max normalization is performed on this to scale the values between -1 to 1. The re-scaled signal is denoted by $\tilde{Z}$, and then $\tilde{Z}$ is encoded to polar coordinates by the following equation:
\begin{equation}
            r_{i} = \frac{i}{n}, \psi_{i} = arccos(\tilde{z_{i}});
            -1\leq \tilde{z_{i}} \leq 1, \tilde{z_{i}} \in \tilde{Z} 
    \end{equation}
Here, $\psi_{i}$ represents the angular cosine of $\tilde{z_{i}}$, (\textit{i=1,2, ..., n}) that corresponds to the angle in the polar coordinate system and $r_{i}$ represents the radius of polar coordinate at each timestamp \textit{'i'}. As $\tilde{Z}$ ranges in [-1, 1], $\psi_{i}$ ranges in angular bounds [0, $\pi$]. Thus, each time instant is encoded into the radius of the polar coordinates and the changes in amplitude of the signal are encoded into the angle of the polar coordinates. This bijective mapping from 1D signal to 2D space is referred as GAF that ensures no information loss. GAF can generate two types of images by calculating the sum and difference of $\psi_{i}$ between all pairs of sampling points. These are (i) Gramian Angular Summation Field (GASF) and (ii) Gramian Angular Difference Field (GADF). The computation of GASF and GADF are shown is Equation~\ref{eq:gasf} and ~\ref{eq:gadf}, respectively.

\begin{equation}
\label{eq:gasf}
    GASF = [cos(\psi_{i}+\psi_{j})]_{N \times N},  i,j = 1,2,..., N
\end{equation}

\begin{equation}
\label{eq:gadf}
    GADF = [sin(\psi_{i}-\psi_{j})]_{N \times N},  i,j = 1,2,..., N
\end{equation}
Thus, in GAF matrix, each element is the cosine of the summation (GASF) or sine of the difference (GADF) of pairwise temporal values. The visualization of GASF and GADF of fMRI TS of COCO, Imagenet and SUN of a specific active voxel is shown in the first and second row of Figure~\ref{fig:2dimages}, respectively.

\subsubsection{Markov Transition Field}
\begin{figure}[htbp]
\centering
\includegraphics[width=6.5cm]{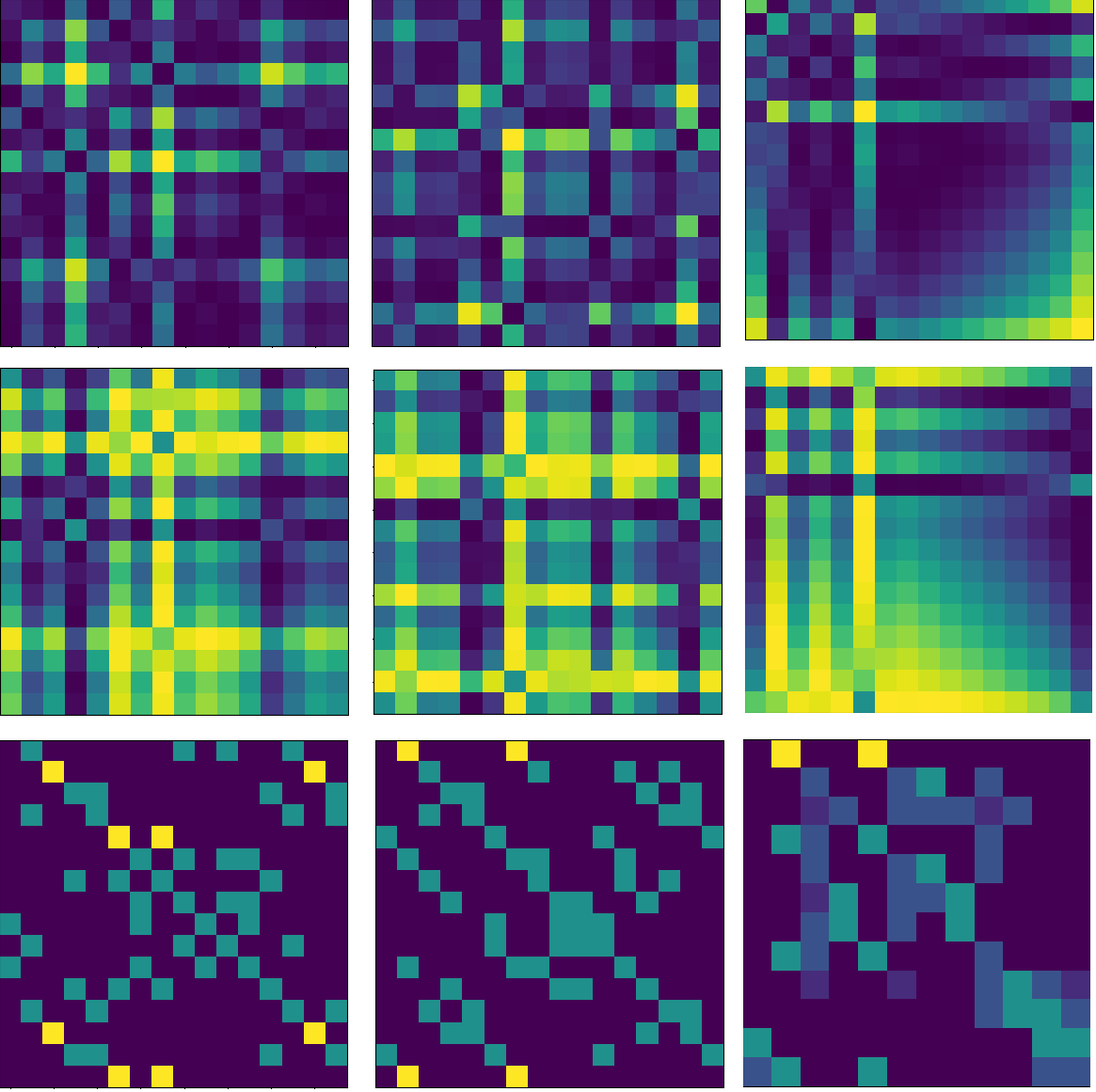} 
\caption{Illustration of 2D representation of BOLD fMRI TS as obtained using GAF and MTF across COCO (\textit{first column}), Imagenet (\textit{second column}) and SUN (\textit{last column}). The rows in the figure denotes 2D representation of GASF (\textit{top row}), GADF (\textit{second row}) and MTF (\textit{last row}).}
\label{fig:2dimages}
\end{figure}

Different from GAF, spatial encoding of 1D TS is obtained using MTF from the transition probabilities between adjacent pairs of elements in the 1D data sequence. 
Let us consider fMRI time series as $Z=\{z_{1}, z_{2}, z_{3}, \dots, z_{n}\}$. First, the values of the time series is quantize into Q quantile bins, where each value $Z_{i}$ is mapped to its corresponding $q_{j} (j\in [1, Q])$. 
Then a $Q \times Q$ weighted adjacency matrix 'W' is constructed where each element $w_{ij} (1\leq i, j\leq Q)$ in W is the frequency (or counts) with which a data point in the state $q_{j}$ is followed by a data point in the state $q_{i}$. This adjacency matrix 'W' is known as Markov transition matrix. It is shown is Equation~\ref{eq:mtf1}.
After normalization with $\sum_{j}w_{ij}=1$, W becomes the MTF (Equation~\ref{eq:mtf2}), in which each element in this matrix can be regarded as the probability that a data point in the state $q_{i}$, transiting to the state $q_{j}$ ($q_{i} \rightarrow q_{j}$). Thus, at each pixel location, transition probability from the quantile at time step 'i' to the quantile at time step 'j' is assigned which actually encodes multi-step transition probabilities of the TS. This in-turn captures the transition dynamics of the TS between different time lags. The visualization of 2D MTF, as obtained from fMRI TS of COCO, Imagenet and SUN of a specific active voxel is shown in the last row of Figure~\ref{fig:2dimages}.

\begin{equation}
\label{eq:mtf1}
    W =
    \begin{bmatrix}
    w_{11|P(x_{t}\in q_{1}|x_{t-1} \in q_{1})} & \cdots & w_{1Q|P(x_{t}\in q_{1}|x_{t-1} \in q_{Q})} \\
    \vdots & \ddots & \vdots \\
    w_{i1|P(x_{t}\in q_{i}|x_{t-1} \in q_{1})} & \cdots & w_{iQ|P(x_{t}\in q_{i}|x_{t-1} \in q_{Q})} \\
    \vdots & \ddots & \vdots \\
    w_{Q1|P(x_{t}\in q_{Q}|x_{t-1} \in q_{1})} & \cdots & w_{QQ|P(x_{t}\in q_{Q}|x_{t-1} \in q_{Q})}
    \end{bmatrix}
\end{equation}

\begin{equation}
\label{eq:mtf2}
    MTF =
    \begin{bmatrix}
    w_{ij|x_{1}\in q_{i},x_{1} \in q_{j}} & \cdots & w_{ij|x_{1}\in q_{i},x_{n} \in q_{j})} \\
    \vdots & \ddots & \vdots \\
    w_{ij|x_{k}\in q_{i}|x_{k} \in q_{j}} & \cdots & w_{ij|x_{k}\in q_{i}|x_{n} \in q_{j}} \\
    \vdots & \ddots & \vdots \\
     w_{ij|x_{n}\in q_{i}|x_{1} \in q_{j}} & \cdots & w_{ij|x_{n}\in q_{i}|x_{n} \in q_{j}} \\
    \end{bmatrix}
\end{equation}

\subsection{Neural Network Architecture for Classification}

Different types of neural network architectural choices for categorization of images across considered visual datasets using fMRI TS are described in this section. Along with single CNN model, parallel CNN model is also employed to combine 2D spatial features, as obtained from GAF and MTF. Subsequently, LSTM and Bi-LSTM networks are also used to obtain 1D temporal features from fMRI TS.

\subsubsection{\textbf{Input Representation}}
In 2D approach, series of 2D images of size $m\times m\times 1\times n$ are generated from each active voxel-specific fMRI TS, representing COCO, Imagenet and SUN datasets, using GAF and MTF. This is given as input to CNN models, where 'n' refers to the total number of TS generated (number of active voxels) for a specific subject and 'm' is the size of 2D images as generated using GAF and MTF. In 1D approach, the raw fMRI TS of size $m \times 1 \times n$ is directly fed as input to LSTM/Bi-LSTM model.

\subsubsection{\textbf{Single and Parallel CNN Architecture for 2D Representation of fMRI Time Series}}

\begin{figure*}[htbp]
\centering
\includegraphics[width=17cm]{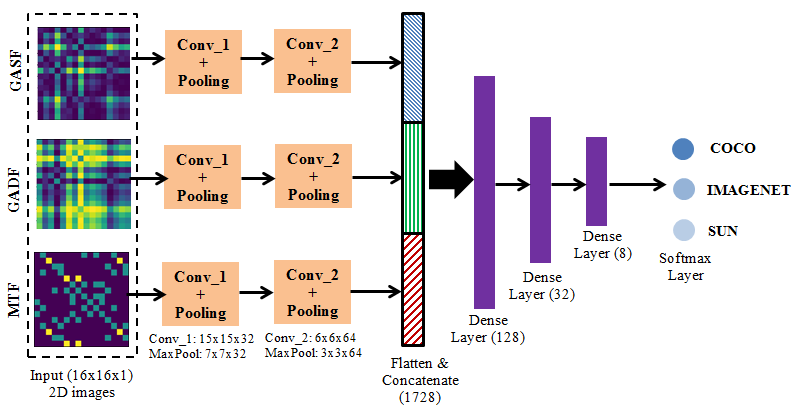} 
\caption{Parallel CNN architecture for classification}
\label{fig:pcnn}
\end{figure*}

Convolutional neural network (CNN) model is used for classification of 2D representation of fMRI TS, as obtained by GAF and MTF, into corresponding dataset of COCO, Imagenet and SUN. Two types of CNN network connections are used in this work (i) single CNN for classification using GAF and MTF, individually and (ii) parallel CNN model that combine spatial features from GASF, GADF and MTF. The single and parallel CNN network models are shown in Figure~\ref{fig:cnnlstm}(\textit{B}) and Figure~\ref{fig:pcnn}, respectively. Single CNN architecture comprises of a series of two convolutional layer of kernel size 32 and 64 respectively and max-pool layer followed by three dense layers of size 576 units, 32 units and 8 units. The same convolutional and max-pooling layers are used in parallel CNN architecture. The features from max-pool layers as obtained from three parallel connections are then concatenated with dense layer of size 1728. Three dense layers of size 128, 32, 8 are added further with softmax (for ternary classification)/sigmoid (for binary classification) activation function in last layer. In all other layers 'ReLU' activation function is utilized.

\subsubsection{\textbf{LSTM and Bi-LSTM for 1D fMRI Time Series}}

\begin{figure}[htbp]
\centering
\includegraphics[width=8.5cm]{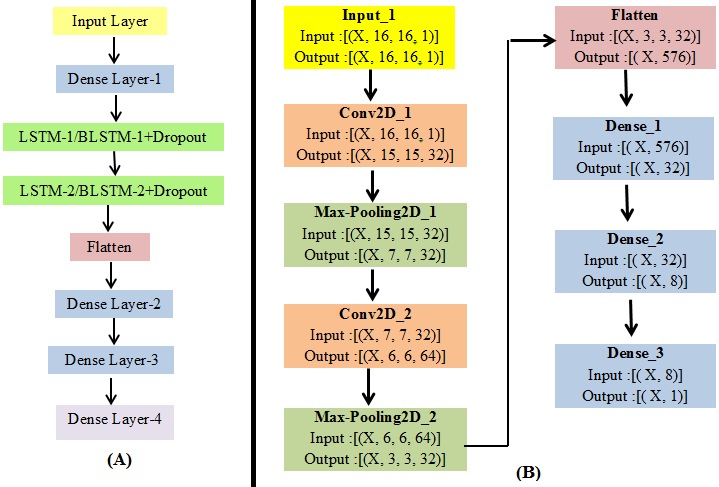} 
\caption{Architecture of (\textit{A}) LSTM and (\textit{B}) CNN model as used in this study.}
\label{fig:cnnlstm}
\end{figure}

In order to compare the classification performance of 2D GAF and MTF with 1D fMRI TS signal, Long Short-Term Memory (LSTM) and Bi-directional LSTM (Bi-LSTM) neural network models are used. In this case, the 1D fMRI TS representing COCO, Imagenet and SUN, is fed as input to LSTM/Bi-LSTM. In both models, two LSTM and Bi-LSTM layers are used. The input TS is fed to first dense layer of size 32 units. The first LSTM, having size 64 units (for Bi-LSTM, it is 128 units) is then applied on this dense layer, followed by a dropout layer (dropout value 0.5) in order to prevent overfitting. Similar LSTM with 32 units (for Bi-LSTM, it is 64 units) and drop out layer are added further and flattened which is followed by two dense layers of size 64 units and 32 units respectively. The fourth dense layer of size 3 units is added at the end for final prediction.'ReLU' activation function is used in all dense layers except the last layer where softmax/sigmoid activation function is applied for final prediction. The LSTM/Bi-LSTM network architecture are shown in Figure~\ref{fig:cnnlstm}(\textit{A})

\subsubsection{\textbf{Training}}

In all deep learning models- CNNs, LSTM and Bi-LSTM network, binary cross-entropy loss (for binary classification) and categorical cross-entropy loss (for multi-class classification) with Adam optimizer is used for training the neural network models. The learning rate is set to 0.001. Batch size is kept as 8 for CNNs, whereas in case of LSTM/Bi-LSTM batch size is kept as 20. Stratified k-fold (\textit{k=10}) cross validation is performed. Colab notebooks and Keras are used to conduct the entire experiment. Colab provides 16GB of NVIDIA Tesla T4 GPU. 

\begin{figure}[htbp]
\centering
\includegraphics[width=9cm]{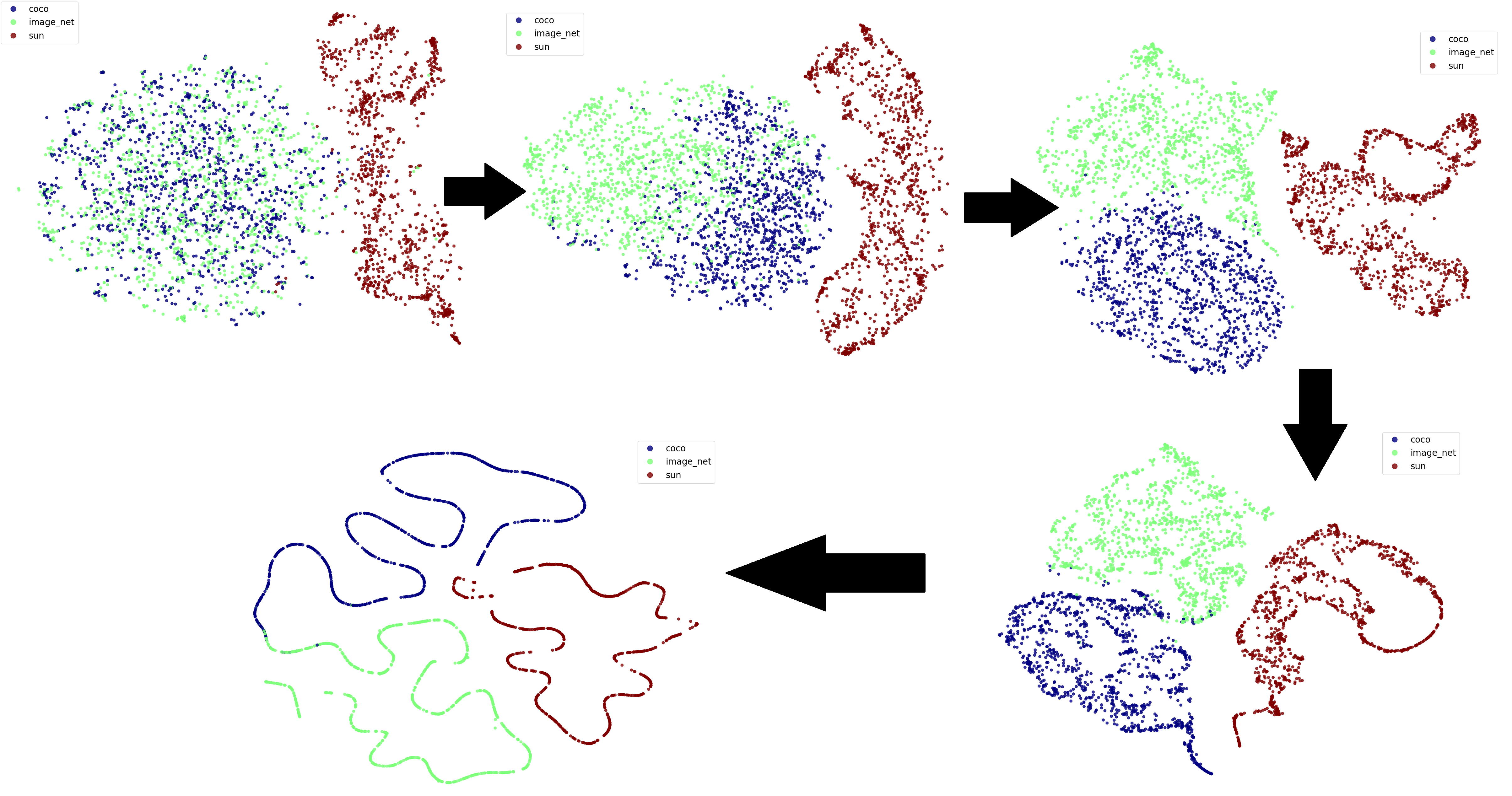} 
\caption{T-SNE visualization of CNN features from flatten to softmax layer in parallel CNN architecture, showing the well-separability of 2D spatial features as obtained using GAF and MTF, in order to classify COCO (\textit{blue}), Imagenet (\textit{green}) and SUN (\textit{red}).}
\label{fig:tsne}
\end{figure}

\begin{table*}[htbp]
\caption{Mean classification accuracy as obtained by 1D and 2D fMRI TS features using deep neural networks across all subjects.}
\label{table:classificationCNN}
\centering
\begin{tabular}{|c|c|c|c|c|c|c|}
\hline
Dim. &Feature &Model & 3-class  &ImageNet  &ImageNet  &COCO  \\
 &  &  &  &Vs. SUN  &Vs. COCO  &Vs. SUN  \\
\hline
1D &Raw &LSTM &0.75 &0.98 &0.63 &0.98 \\
\cline{3-7}
 &fMRI TS  &Bi-LSTM &0.76 &0.99 &0.64 &0.99 \\
\hline
 &MTF &Single &0.87  &0.99  &0.71  &0.99 \\
\cline{2-2}
\cline{4-7}
2D &GASF &CNN &0.87 &0.99 &0.84 &0.98 \\
\cline{2-2}
\cline{4-7}
 &GADF & &0.85 &0.99 &0.86 &0.99 \\
\cline{2-7}
 &MTF+GAF &Parallel CNN &\textbf{0.94} &\textbf{0.99} &\textbf{0.88} &\textbf{0.98} \\
\hline
\end{tabular}
\end{table*}

\section{Results and Discussion}

In order to capture the spatial and temporal information, 2D representation of fMRI TS is obtained using GAF and MTF. Figure~\ref{fig:tsne} shows the t-distributed stochastic neighbor embedding (t-SNE) visualization of CNN features across different layers. The t-SNE analysis is commonly used in deep neural networks for high-dimensional feature visualization in 2D space in order to understand the efficacy of model. The t-SNE plots clearly depicts the well separability of Imagenet, COCO and SUN features, as obtained by GAF and MTF. The performance of classification is tabulated in Table~\ref{table:classificationCNN}. In case of 2D representation of fMRI TS, using 10-fold cross-validation, the highest accuracy of 94\% is obtained for 3-class classification using parallel CNN, across all subjects. Using single CNN model, the 3-class classification using GADF yields 85\% accuracy across subjects. This accuracy is increased to 87\% when MTF and GASF features are used. It is seen that, in case of binary classification, the accuracy is improved drastically to 99\% when classification is performed with respect to SUN. This could be attributed to the fact that the images contain in SUN dataset is very distinct in terms of context and complexity from the images contain in COCO and ImageNet. However, in classifying Imagenet and COCO this accuracy is reduced to 88\%, which could be due to the similarities in spatial context between the images of these two datasets. One example of such spatial context similarity is shown in Figure~\ref{fig:imgs}: the baseball ground image in COCO and ImageNet that contain similar information, leading to high probability of misclassification. The result of GAF and MTF is also compared with classification using 1D raw fMRI TS data. LSTM and Bi-LSTM models are used for classification using 1D TS data. Here, although the 2-class classification results with respect to SUN yield similar accuracy as seen in case of 2D, the average accuracy of 3-class is decreased to 18\%. Moreover, using 1D TS data the 2-class classification between Imagenet and COCO gives only 64\% accuracy, which is improved drastically by 22\% when 2D representation of fMRI TS is utilized for classification. \par
Vision science, particularly machine vision, has been revolutionized by introducing large-scale image datasets and statistical learning approaches. However, human neuroimaging studies of visual perception is still remain the fundamental open problem. The main motivation of this study is to investigate - does our brain function differently when we look at an image of a lion versus a mountain with different scales and complexities? There are very limited literature available that have addressed this issue. Using fMRI TS, it is possible to quantify the differences in brain activity in response to the visual stimuli, conveying different information. The publicly available “BOLD5000” dataset \cite{b1}, containing fMRI images, acquired during visual tasks, provides a great opportunity to understand differences in brain functional activities related to vision. The objective of this study is to classify the visual stimuli having diverse image complexities into corresponding visual datasets of Imagenet, COCO and SUN using fMRI TS. It is evident that the images across these dataset convey different visual information. For example, Imagenet images have singular-focused unoccluded object, occupying large image fraction and illumination uniformity. COCO images contain single/multiple objects, seen in everyday life, at varying scales. SUN contains images of natural settings with cluttered background, changes in illumination and occlusions, which do not focus on any specific object(s). These differences are examined in recent literature \cite{b16,b17}. In this study, we have identified these differences by categorizing fMRI TS of visual stimuli into corresponding visual datasets. As discussed in Section~\ref{sec:intro}, comparison across visual datasets is essential in visual neuroscience research, since it reveals how neural activity results in visual perception. 
To address this, spatial encoding of fMRI BOLD TS using GAF and MTF has been performed to identify the latent pattern, if distinct, across images of Imagenet, COCO and SUN. For classification purpose, 2D CNN architecture is utilized. The classification performance of 2D CNN is also compared with 1D neural network models- LSTM and Bi-LSTM where the raw 1D fMRI TS is used for classification. It is found that classification of images across visual datasets using 2D fMRI TS outperforms classification using 1D TS data; especially the result of 3-class classification and classification of images of COCO vs. Imagenet improves substantially, when parallel CNN model with combined 2D GAF and MTF features are utilized. \par
To the best of knowledge of authors, this is the first study that utilizes 1D and 2D representation of BOLD fMRI TS to categorize diverse image complexities between distinct datasets. The proposed methodology outperforms previous work by A. Jamalian et.al. \cite{b2} on BOLD5000 dataset in which authors used sequence models to classify only three well-separable classes of images: animal, artifact and scenes and achieved the accuracy of 68\%. There are two more studies reported in literature that used BOLD5000 dataset for other tasks. A. Jaiswal et.al.\cite{b3} used BOLD5000 images to pretrain deep neural network models for predicting cognitive fatigue in traumatic brain injury. Subba R. Oota et.al.\cite{b4} used BOLD5000 dataset to study brain encoding models that aims to reconstruct fMRI brain activity given a stimulus. Contrary to these studies, the current work presents a different approach that focused on analysing the efficacy of utilizing 2D representation of fMRI TS to categorize images with different complexity across visual datasets.
In future, the present work can be extended to investigate correlation between these fMRI BOLD TS from which visual brain network can be constructed for COCO, Imagenet and SUN. Graph-theoretical analysis can be performed to check the differences in topological architecture of these visual brain networks which could further clarify how differently human brain reacts while looking into images with varied complexities. 

\section{CONCLUSIONS}

The work presents fMRI TS analysis to categorize distinct image complexities across COCO, ImageNet and SUN. However, BOLD5000 data contains only 4 subjects, which is a major limitation towards conclusive inferences. Thus, generalization of this study require more number of participants, fMRI sessions and diversity of stimulated images, that will help advance understanding of human visual functional neural networks. Although 5,254 images with diverse complexity are quite large to study brain visual dynamics using fMRI, it is still relatively smaller when compared to human visual experience in everyday life. Nevertheless, as a baseline work, the results showed a good foundation for future fMRI studies on how vision is represented in brain.  

\addtolength{\textheight}{-12cm}   


\section*{ACKNOWLEDGMENT}

The authors would like to thank Mphasis F1 Foundation, Cognitive Computing grant to conduct research at IIITB.


\begin{thebibliography}{99}

\bibitem{b1} Chang, Nadine, John A. Pyles et.al. "BOLD5000, a public fMRI dataset while viewing 5000 visual images." Scientific data 6, no. 1: 1-18, 2019.
\bibitem{b2} A. Jamalian, R. Ayub, F. Fadavi. Categorization of seen images from brain activity using sequence models. Article, CS230: Deep Learning, Stanford University, CA, 2019.
\bibitem{b3} Jaiswal, Ashish, Ashwin Ramesh Babu et.al. Understanding Cognitive Fatigue from fMRI Scans with Self-supervised Learning, arXiv:2106.15009, 2021.
\bibitem{b4} Subba, Reddy Oota, Jashn Arora et.al. Visio-Linguistic Brain Encoding. arXiv-2204, 2022.
\bibitem{b5} ZhiguangWang and Tim Oates. Imaging Time-Series to Improve Classification and Imputation. arXiv:1506.00327v1, 2015.
\bibitem{b6} Wang, Z. and Oates, T. Encoding time series as images for visual inspection and classification using tiled convolutional neural networks. In Proc. Workshops at the Twenty-Ninth AAAI Conference on Artificial Intelligence, Austin, TX, USA, pp:25–30, 2015.
\bibitem{b7} Wang M, Wang W, Zhang X, Iu HH-C. A New Fault Diagnosis of Rolling Bearing Based on Markov Transition Field and CNN. Entropy. 24(6):751, 2022. 
\bibitem{b8} Chunli Lei, Linlin Xue et.al. Rolling bearing fault diagnosis by Markov transition field and multi-dimension convolutional neural network. Meas. Sci. Technol. vol. 33:114009, 2022.
\bibitem{b9} K. Thanaraj, B. Parvathavarthini et. al. Implementation of Deep Neural Networks to Classify EEG Signals using Gramian Angular Summation Field for Epilepsy Diagnosis. ArXiV, vol. abs/2003.04534, 2020.
\bibitem{b10} A. Shankar, S. Dandapat and S. Barma. Seizure Type Classification Using EEG Based on Gramian Angular Field Transformation and Deep Learning. In Proc. International Conference of the IEEE Engineering in Medicine and Biology Society (EMBC), pp. 3340-3343, 2021
\bibitem{b11} Abouzar E. and Roozbeh R. Single Residential Load Forecasting Using Deep Learning and Image Encoding Techniques. Electronics. 9, 68, 2020.
\bibitem{b12} Jehn-Ruey Jiang and Cheng-Tai Yen. Markov Transition Field and C onvolutional Long Short-Term Memory Neural Network for Manufacturing Quality Prediction. In Proc. IEEE International Conference on Consumer Electronics-Taiwan (ICCE-Taiwan), 2020.
\bibitem{b13} Wei Sun, Jie Zhou et. al. Markov Transition Field Enhanced Deep Domain Adaptation Network for Milling Tool Condition Monitoring. Micromachines 2022, 13, 873, 2022.
\bibitem{b14} S. Xia, Z. Pan, Z. et.al. Malware Classification with Markov Transition Field Encoded Images. Eighth International Conference on Instrumentation and Measurement, Computer, Communication and Control (IMCCC), pp. 1-5, 2018.
\bibitem{b15} Velasco GC and L. Iraklis. Analysis of Time Series Imaging Approaches for the Application of Fault Classification of Marine Systems. In Proc. 32nd European Safety and Reliability Conference (ESREL 2022), 2022.
\bibitem{b16} K. Gauen et al. Comparison of Visual Datasets for Machine Learning. In Proc. IEEE International Conference on Information Reuse and Integration (IRI), San Diego, USA, pp. 346-355, 2017.
\bibitem{b17} Bilal Alsallakh, Pamela Bhattacharya et.al. A Tour of Visualization Techniques for Computer Vision Datasets. arXiv:2204.08601v1, 2022.


\end{thebibliography}
\end{document}